\documentclass[amsmath,twocolumn,prl,aps]{revtex4-1}
\usepackage{amsthm,amsfonts,graphicx,verbatim,color}
\usepackage{bm}
\usepackage[T1]{fontenc}
\usepackage[utf8]{inputenc}
\usepackage[colorlinks=true, linkcolor = blue, citecolor = blue, urlcolor = blue]{hyperref}
\usepackage{lmodern}
\usepackage{enumitem}

\newcommand{\mb}[1]{\ensuremath{\mathbf #1}}
\newcommand{\mc}[1]{\ensuremath{\mathcal #1}}

\newcommand{\ket}[1]{|#1\rangle}
\newcommand{\eq}[1]{Eq.~\eqref{eq:#1}}
\newcommand{\fig}[1]{Fig.~\ref{fig:#1}}
\newcommand{\angles}{\ensuremath{\boldsymbol \theta}}
\newcommand{\critexp}{\nu_{2\text{\fontsize{4.5}{4}\selectfont D}}}
\newcommand{\rhotop}{\rho_{\rm top}}

\begin{document}

\title{Dimensional crossover of the integer quantum Hall plateau transition and disordered topological pumping}

\author{Matteo Ippoliti${}^1$ and R. N. Bhatt${}^2$}
\affiliation{${}^1$Department of Physics and ${}^2$Department of Electrical Engineering, Princeton University, Princeton NJ 08544, USA}

\begin{abstract}
We study the quantum Hall plateau transition on rectangular tori.
As the aspect ratio of the torus is increased, the two-dimensional critical behavior, 
characterized by a subthermodynamic number of topological states in a vanishing energy window around a critical energy,
changes drastically. 
In the thin-torus limit, the entire spectrum is Anderson-localized;
however, an \emph{extensive} number of states retain a Chern number $C\neq 0$. 
We resolve this apparent paradox by mapping the thin-torus quantum Hall system onto a disordered Thouless pump, where the Chern number corresponds to the winding number of an electron's path in real space during a pump cycle.
We then characterize quantitatively the crossover between the one- and two-dimensional regimes for finite torus thickness, where the average Thouless conductance also shows anomalous scaling.
\end{abstract}

\maketitle


\textit{Introduction.}
The integer quantum Hall plateau transition~\cite{VonKlitzing1980} has a long and rich history as an example of the interplay between disorder and topology in condensed matter. 
While the quantization is ultimately due to the presence of a topological invariant~\cite{Laughlin1983, Thouless1982}, its astonishing precision is due to disorder-induced localization of electron states away from the critical energy~\cite{Prange1981}.
In a high magnetic field, the motion of electrons is confined to the lowest Landau level (LLL). 
The LLL carries a non-zero \emph{Chern number}, a topological invariant related to the Hall conductance, which forbids complete localization of the spectrum. 
A critical energy exists where the electron localization length $\xi$ diverges, explaining the plateau transition as a quantum critical point that has successfully been studied by means of scaling theories~\cite{Huckestein1995}.
However, the precise value of $\nu$, the critical exponent characterizing the divergence of $\xi$, and whether or not it agrees with experiment~\cite{Wei1988, Engel1993, Li2009, Kivelson1992}, remains controversial~\cite{Gruzberg2017, Zhu2019, Puschmann2019}.

Most numerical studies of the critical exponent have relied on the transfer matrix method for either the original continuum LLL problem~\cite{Huckestein1990, Huckestein1992} or the Chalker-Coddington network model~\cite{Chalker1988, Lee1993, Slevin2009, Obuse2012} on strip geometries.
On the other hand, purely two-dimensional methods to determine $\nu$ have been developed based on the topological character of individual eigenstates~\cite{Huo1992, Arovas1988} (an idea that has since been used in several studies~\cite{Yang1996, Yang1997, Yang1999, Sheng1995, Sheng1997, Sheng2003, Wan2005, Zhu2019}), the disorder-averaged Hall~\cite{Bhatt1991, Ippoliti2018}, Thouless~\cite{Priest2014, Ippoliti2018} and longitudinal~\cite{Soukoulis1998} conductance,
as well as quantum diffusion~\cite{Sandler2004}.
Here one considers a square torus with both sides scaled concurrently, $L_x = L_y \sim N_\phi^{1/2}$ ($N_\phi$ is the number of magnetic flux quanta through the system, proportional to the system's area). 
The number of states with nonzero Chern number (hereafter simply called Chern states) is found to diverge subextensively with system size, as $N_\phi^{1-\frac{1}{2\nu}}$~\cite{Huo1992}.
The success of methods based on the Chern number in square geometry motivates their application to rectangular geometries $L_x > L_y$ with varying aspect ratio $a = L_x/L_y$, and particularly in the quasi-one-dimensional limit $a\to \infty$ at fixed thickness, reminiscent of the transfer matrix calculations.
This is especially interesting because the defining feature of the 2D problem (the presence of a topologically robust Hall conductance, encoded in the {Chern number} $C$) does not have an obvious one-dimensional counterpart. 
While the mathematical definition of $C$ holds regardless of system size or aspect ratio,
on physical grounds the system in the quasi-1D limit must be described by a local, disordered free-fermion chain -- essentially the Anderson model~\cite{Anderson1958}. 
This raises the question of what happens to Chern states in this limit, and how the topological character of the LLL is manifested once the system is mapped onto a 1D Anderson insulator.

One may reasonably expect, given the stronger tendency towards localization in one-dimensional systems~\cite{Abrahams1979}, that quasi-1D scaling will lead to a faster decay of the fraction of Chern states relative to the 2D case (where the fraction falls off as $N_\phi^{-\frac{1}{2\nu}}$), 
perhaps even  to saturate the lower bound $N_\phi^{-1}$ 
(achieved if all states but one have $C=0$).
In fact, we find quite the opposite:
Chern states do \emph{not}  {vanish} under 1D scaling. 
On the contrary, they represent a {finite} fraction of all states -- and asymptotically take over the entire spectrum! 

As a byproduct, we also obtain the (longitudinal) Thouless conductance $g$~\cite{Thouless1974}. 
Both the typical and average $g$ decay exponentially with $L_x$, as is expected for localized one-dimensional systems.
Interestingly though, we find that the \emph{average} $g$ retains a memory of the 2D critical scaling.

Existing studies of one-dimensional scaling of the integer quantum Hall problem~\cite{Ketteman2004, Struck2005} focus on open boundary conditions, where the crossover is seen through mixing of topological edge states on opposite edges of the strip.
Our edge-free torus geometry offers a different perspective on the problem and reveals fascinating and unexpected behavior. 
Guided by these surprising numerical findings, we develop a theoretical understanding based on a mapping to a {disordered Thouless pump}~\cite{Thouless1983} and clarify the meaning of the Chern number in the 1D limit. A quantitative description of the proliferation of Chern states follows naturally from this perspective.


\textit{Model and numerical method.} 
We consider a continuum model of two-dimensional (2D) electrons in a high perpendicular magnetic field such that the dynamics can be projected onto the LLL. 
The model is set on a rectangular torus with sides $L_x$, $L_y$ such that $L_x L_y = 2\pi N_\phi \ell_B^2$, where $\ell_B = \sqrt{eB/\hbar}$ is the magnetic length, which we set to 1 henceforth.
We define the aspect ratio $a=L_x/L_y$ and take $a\geq 1$.
Disorder in the system is modeled by a Gaussian white noise potential $V(\mb r)$, $\langle V(\mb r_1) V(\mb r_2)\rangle = U^2 \delta^2(\mb r_1 - \mb r_2)$. 
We set $U=1$ henceforth as disorder is the only energy scale in the problem: kinetic energy is quenched in the LLL; the cyclotron gap and interaction strength are taken to be infinite and zero respectively.
The torus has generalized periodic boundary conditions with angles $\angles_{x,y}$.
These also represent magnetic fluxes through the two nontrivial loops in the torus and are needed to define and compute Chern numbers of individual eigenstates in the disordered problem.
For each disorder realization, we compute and diagonalize the single-particle Hamiltonian on a lattice of boundary angles $\angles$ and store the eigenvalues $\{E_n(\angles)\}$ and eigenvectors $\{ \ket{\psi_n(\angles)} \}$.
The energies are used to calculate the Thouless conductance 
$g_n \equiv \mathbb E_{\theta_y} [ \sigma_{\theta_x} E_n(\angles)]$
($\mathbb E$ denotes averaging, $\sigma$ denotes standard deviation), a measure of sensitivity to boundary conditions in the \emph{long} direction;
the wavefunctions are used to compute each eigenstate's Chern number $C_n$ via a standard numerical technique~[S1]. 
Further details on the model and the numerical method are provided in the Supplemental Material~\footnote{See Online Supplemental Material 
for details on the numerical calculations, a proof of the identity between Chern number and winding number, properties of the electrons' random walks, and additional data on the density of Chern states and electron localization length.}.

\textit{Density of Chern states.}
With the method outlined above, we calculate the density of states with Chern number $C$, $\{ \rho_C(E): C\in \mathbb Z\}$.
These obey $\sum_{C} \rho_C = \rho $ (total density of states) as well as $\sum_{C} C\rho_C = \partial_E \sigma_{xy} $
(Hall conductance). 
Past studies~\cite{Huo1992, Zhu2019} have characterized the 2D critical behavior by looking at the density of ``current-carrying states'', $\rhotop(E) \equiv \rho(E)-\rho_0(E)$.
The width of $\rhotop$ scales as $N_\phi^{-1/2\critexp}$ in the 2D thermodynamic limit. 

\begin{figure}
\centering
\includegraphics[width=\columnwidth]{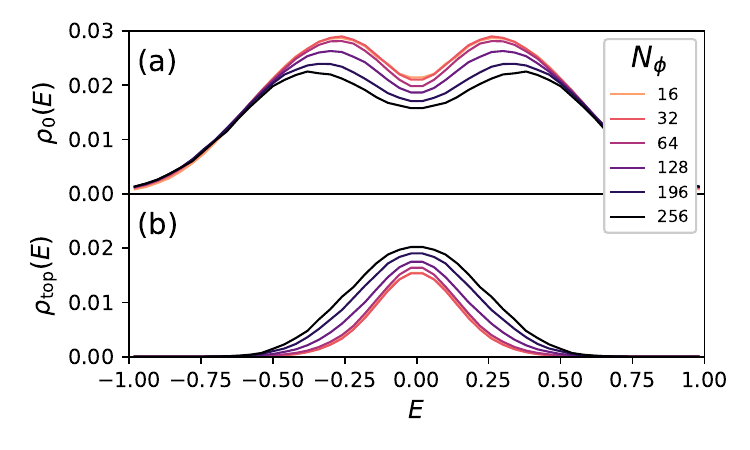}
\caption{Density of (a) $C=0$ and (b) $C\neq 0$ states for fixed $L_y = 10$ and increasing $N_\phi$.
The density of Chern $C\neq 0$ states $\rhotop(E)$ grows and broadens at the expense of $\rho_0(E)$. \label{fig:chern}}
\end{figure}

In the present context, we observe completely different behavior.
Namely, the width of $\rhotop$ does \emph{not} vanish as $L_x$ is increased. 
It stays roughly constant for $a \gtrsim 1$, and eventually starts \emph{increasing} for $a \gg 1$ (Fig.~\ref{fig:chern}).
This increase is due both to the broadening of $\rho_{\pm 1}(E)$ (i.e. more pairs of Chern $\pm 1$ states appearing away from the band center), and to an increase in higher-$|C|$ states. 
Despite these effects, the Hall conductance remains unchanged, and is determined by the shortest side of the torus~\cite{Note1}.
It is as if percolating in \emph{either} direction is enough for a state to acquire a nonzero Chern number.

This extensive number of topological ``current-carrying states'' seems to be incompatible with the localized nature of the spectrum (which we verify independently by means of the Thouless conductance and localization length).
Reconciling these facts requires a careful analysis of the fate of Chern numbers as the dimensionality is tuned from $d=2$ to $d=1$ by increasing the aspect ratio $a$.

\textit{Thin-torus limit.}
The above question is best addressed in the thin-torus limit $L_y \ll 1$, though (as we shall clarify later) the answer we find also applies to finite $L_y$, provided $a$ is sufficiently large.
The LLL Hamiltonian in the thin-torus limit is approximated by 
\begin{equation}
H_{1D} = \sum_n v_n c_n^\dagger c_n + (t_n c_{n+1}^\dagger c_n + h.c.) \;,
\label{eq:1dham}
\end{equation}
with $v_n = V_0(x_n)$, $t_n = e^{i\theta_x/N_\phi} V_1(x_n) $, and $x_n = (2\pi n + \theta_y)/L_y$. 
The $V_m$ are partial Fourier transforms of the LLL-projected real-space disordered potential, $\tilde{V}(x,y)$, given by
\begin{equation}
V_m(x) \equiv \int_0^{L_y} \frac{dy}{L_y} e^{2\pi i m y/L_y} \tilde{V}(x,y) \;.
\label{eq:Vm}
\end{equation}
LLL-projection suppresses non-zero wave vectors, giving $t/v \sim e^{-\pi^2/L_y^2} \ll 1$. 
Further-neighbor hopping terms in \eq{1dham} are exponentially smaller than $t$ and can be neglected.
In the following, we take $t_n \equiv t e^{i\theta_x/N_\phi}$ for simplicity, as the precise magnitudes are unimportant. 
The angles $\angles$ assume very different roles in this asymmetric limit:
$\theta_x$ is the magnetic flux through the ring, while $\theta_y$ is the parameter of a Thouless pump~\cite{Thouless1983} which smoothly moves the Landau orbits relative to the background potential.
At any fixed $\angles$, the Hamiltonian of \eq{1dham} is Anderson localized.
As the pump parameter $\theta_y$ is adiabatically taken through a cycle, the random on-site potentials $v_n(\theta_y)$ change smoothly and the system undergoes spectral flow: 
at the end of the cycle, $v_n(2\pi) = v_{n+1}(0)$, so the initial and final spectra coincide up to a $n\mapsto n+1$ translation.
However, following each eigenstate through the adiabatic cycle reveals an interesting picture. 

Adiabatically changing a local chemical potential in an Anderson insulator leads to non-local charge transport~\cite{Khemani2015} due to avoided resonances between the manipulated site and arbitrarily distant ones (the distance is practically limited by $\log \tau$, where $\tau$ is the time scale of the adiabatic manipulation; for calculating $C$, we can take $\tau \to \infty$).
In the present setting, varying $\theta_y$ adiabatically manipulates \emph{all random fields at once}, giving rise to a complicated network of resonances and thus more intricate patterns of charge transfer across the system.
However, as a consequence of adiabaticity, an electron that starts the cycle in orbital $n$ ends in orbital $n-1$ (i.e., at the same point in real space).
Whenever two sites $n_1$ and $n_2$ are tuned past a resonance, charge is transported by a sequence of virtual nearest-neighbor hops through the shortest path between them~\cite{Note1}.
One may expect each electron to take a local random walk in the vicinity of its initial site $n_i$ before ending the cycle at site $n_i-1$.
However, this cannot be the case for every electron: 
at least one must wind around the entire system.
Simple algebra shows that the winding numbers $W_n$ of the electrons' paths must satisfy $\sum_n W_n=1$~\cite{Note1}.

This bears intriguing similarity to the total Chern number of states in the Landau level, $\sum_n C_n = 1$. 
In fact, such an identification is correct: the Chern number $C_n$ reduces to the winding number $W_n$ in the thin-torus limit. 
This can be seen by considering the phase acquired during a loop around the ``Brillouin zone'' defined by $\angles$. 
Threading flux $\theta_x$ does nothing to an Anderson localized wavefunction, whereas threading a quantum of $\theta_y$ flux causes it to wind $W_n$ times around the ring, which encircles the $\theta_x$ flux. 
The net phase acquired is thus $2\pi W_n$, giving $C_n = W_n$.
This can be straightforwardly made rigorous by partitioning the $\angles$ torus into thin rectangular strips, so phases are defined unambiguously~\cite{Note1}.
This identification is the key to explaining the observed proliferation of Chern states under 1D scaling. 
In essence, during a Thouless pump cycle, every electron hops randomly and non-locally across the chain many times, generically acquiring a large winding number, and thus a large Chern number.
Quantitatively, we find the number of steps in the random walk $N_r$ diverges as $N_r \sim L_x$; 
the distribution of Chern numbers is approximately normal, with standard deviation $\sim L_x^{1/2}$~\cite{Note1}.


\begin{figure}
\centering
\includegraphics[width=\columnwidth]{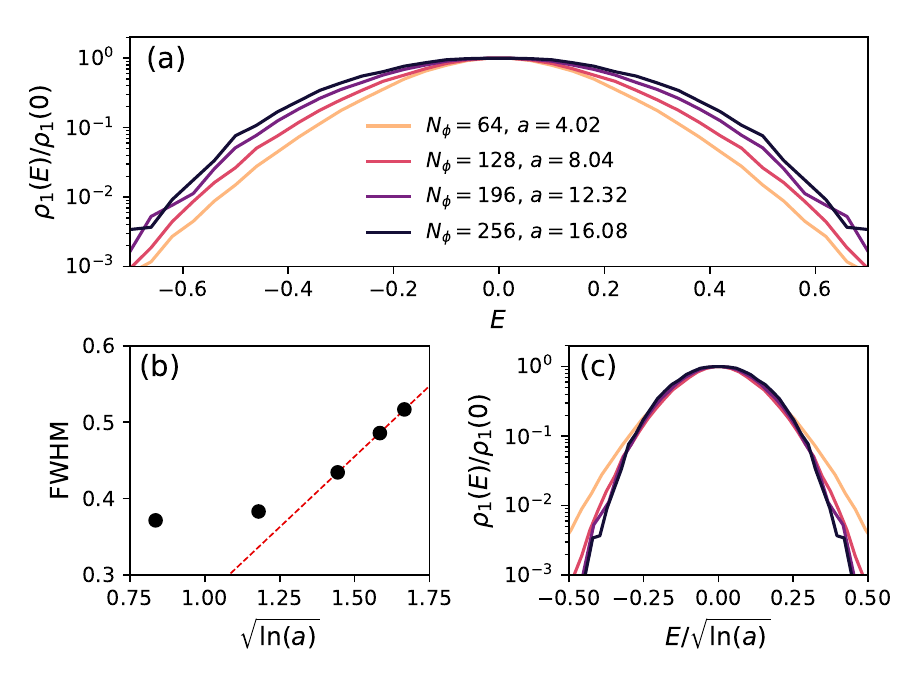}
\caption{(a) Density of $C=1$ states $\rho_1(E)$ for $L_y = 10$ and varying aspect ratio $a$.
(b) Full width at half maximum of $\rho_1$ indicates broadening consistent with \eq{dim_crossover}.
(c) Rescaling the energy by $\sqrt{\ln(a)}$ collapses the curves for different sizes.
\label{fig:aspectratio}}
\end{figure}

\textit{Dimensional crossover.}
Even though the thin-torus limit $L_y\ll 1$ is a helpful simplification, the physics described above remains valid for $L_y>1$, as long as $a \gg 1$. 
Hopping matrix elements are significant up to a real-space distance $\mathcal O(1)$, i.e. a number of sites $\mathcal O(L_y)$. 
These matrix elements are responsible for local level repulsion and strongly suppress energy fluctuations during the Thouless pump cycle. 
On a square torus, we know from numerics that the average Thouless conductance obeys
$ g(E,L) \simeq G(EL^{1/\critexp})$, where $G(x) \simeq g_0 e^{-x^2/2\sigma^2}$ is a scaling function and $g_0$ and $\sigma$ are $\mc O(1)$ constants.
Inverting the definition of $g$ yields an estimate of the energy fluctuation $\delta E$ of a typical state during the pump cycle:
\begin{equation}
\delta E \sim 
\frac{2\pi vg_0}{L^2} \exp\left(-\frac{E^2}{2\sigma^2} L^{2/\critexp}\right)\;.
\label{eq:deltaE}
\end{equation}
Here $v$ is the bandwidth and $2\pi v/L^2$ is the typical level spacing.
As $\delta E$ is determined by the range of local hopping matrix elements,
\eq{deltaE} remains true if we consider a rectangular torus and replace $L$ with the \emph{short} circumference $L_y$. 
The expected number of resonances encountered during a pump cycle, $N_r$, is proportional to the number of states in the spectrum with energies within the range of fluctuations $\delta E$. 
Approximating $\rho(E)\simeq \frac{L_xL_y}{2\pi v} e^{-\frac{1}{2}(E/\sigma')^2} $ (the exact expression~\cite{Wegner1983} deviates slightly from a Gaussian) yields
\begin{align}
N_r & \sim  \rho(E) \delta E \sim g_0 a e^{ -\frac{1}{2} (E/E_0)^2 } \;,
\label{eq:Nr}
\end{align}
where $E_0$ is an $L_x$-independent energy scale.
Thus, even away from the band center, and even for $L_y > 1$, increasing $a$ eventually leads to $N_r \gtrsim 1$.
At that point the crossover between 2D and 1D behavior takes place, with typical states acquiring nontrivial winding and thus Chern number.
This crossover happens unevenly in energy:
it starts at the band center (where one already has Chern states even in the 2D thermodynamic limit) and spreads towards the band edges.
The contour defining the crossover (fixed by setting $N_r \simeq 1$) is
\begin{equation}
E \sim \sqrt{\ln (a)} \;.
\label{eq:dim_crossover}
\end{equation}
This prediction is borne out by numerical data on the density of Chern states, $\rho_C(E)$. 
\fig{aspectratio} shows that the broadening of $\rhotop$, already visible in \fig{chern}, is explained fairly accurately as a scaling collapse of $\rho_C(E / \sqrt{\ln(a)})$, for large enough $a$.


\begin{figure}
\centering
\includegraphics[width=\columnwidth]{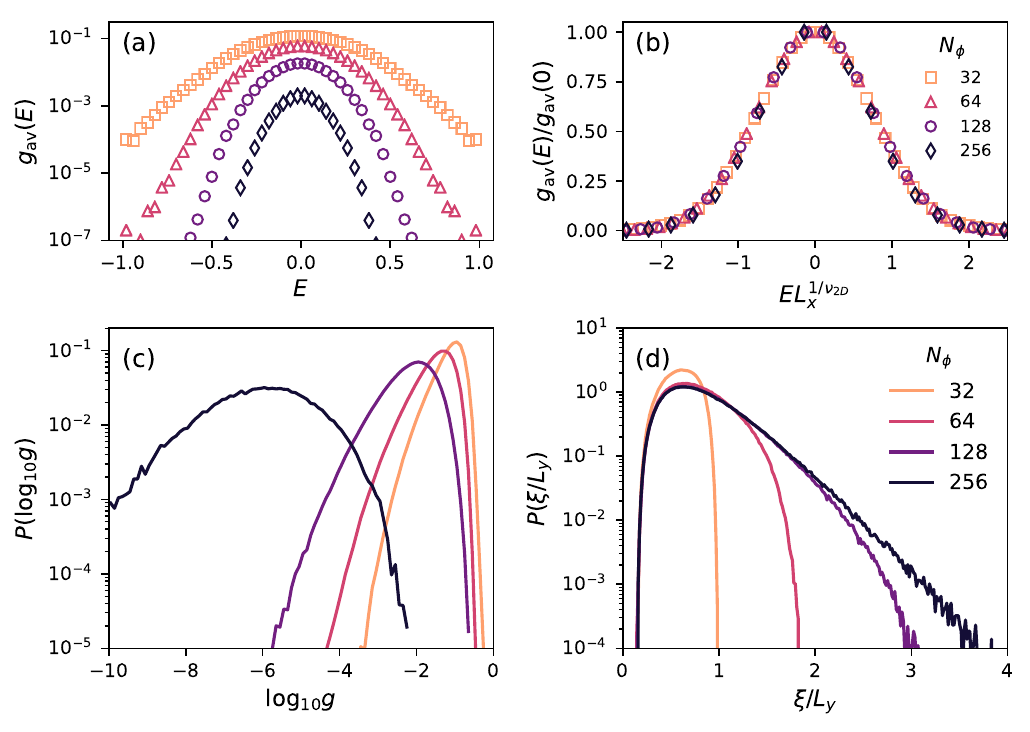}
\caption{
(a) The average Thouless conductance $g_{\rm av}(E)$ for $L_y=14$ decays with increasing $N_\phi$.
(b) The normalized quantity $g_{\rm av} (E)/g_{\rm av}(0)$ shows scaling collapse under $E\mapsto E L_x^{1/\critexp}$ with critical exponent $\critexp \simeq 2.4$.
(c) Distribution of $\log_{10}g(0)$ for $L_y=10$, increasing $N_\phi$.
(d) Distribution of localization length $\xi$ for the same system sizes. An exponential tail $P\sim e^{-c\xi/L_y}$ develops as $N_\phi$ increases.
\label{fig:thouless}}
\end{figure}

\textit{Thouless conductance.}
As mentioned earlier, while 1D scaling causes the proliferation of Chern states across the spectrum, it also removes the critical energy characteristic of the 2D problem and makes the entire spectrum Anderson-localized.
We verify this numerically by calculating the disorder- and eigenstate-averaged Thouless conductance $g_{\rm av} (E)$, Fig.~\ref{fig:thouless}(a).
Unlike the 2D case, where at the center of the band $g_{\rm av}(0) \sim \mc O(1)$ as $L\to\infty$, here we have $g_{\rm av}(0) \sim e^{-L_x/\xi_1}$, as expected for a 1D problem (we find $\xi_1\simeq 1.7L_y$).
However, surprisingly, the normalized quantity $g_{\rm av} (E)/g_{\rm av} (0)$ displays scaling collapse with the \emph{same} critical exponent as the two-dimensional case, $\critexp \sim 2.4$~\footnote{Such collapse is not seen for typical $g$.}, Fig.~\ref{fig:thouless}(b).
These results seem contradictory:
on the one hand, a finite $\xi_1$ suggests localization across the spectrum with no critical energy; 
on the other, we observe signatures of a divergent $\xi_2 \sim E^{-\critexp}$, reproducing the 2D critical behavior, even as the scaling is purely one-dimensional.

The variation of $g$ across samples and eigenstates sheds light on this issue. 
At the center of the band, the distribution of $g$ broadens as $L_x$ is increased and becomes approximately log-normal (the distribution $P(\ln g)$ is shown in Fig.~\ref{fig:thouless}(c)).
States in the positive tail of the distribution, which are abnormally extended in the \emph{long} direction, dominate the average $g_{\rm av}$.
The appearance of $\critexp$ is to be expected as a consequence of such states:
as they percolate across the sample in $L_x$ but not in $L_y$, they are unaware of the aspect ratio, and thus display the 2D critical behavior. 
However they are exponentially rare, which explains the vanishing amplitude of the signal and its presence in $g_{\rm av}$ but not $g_{\rm typ}$.
An exponential tail in the distribution of electron localization lengths $P(\xi/L_y)$ can be seen in Fig.~\ref{fig:thouless}(d); details on the definition and calculation of $\xi$, as well as additional data, are provided in~\cite{Note1}.


\textit{Discussion.}
We have investigated the fate of the quantum Hall plateau transition when the thermodynamic limit is taken in one dimension only. 
Through numerical diagonalization, we have uncovered surprising and counter-intuitive behavior: 
Anderson localization across the spectrum, accompanied by the proliferation of Chern states.
This led us to investigate the fate of the Chern number, a two-dimensional topological invariant, in the quasi-one-dimensional limit defined by $a = L_x/L_y \gg 1$.
In the thin-torus limit $L_y\ll 1$, the system maps onto a 1D Anderson model with a Thouless pump parameter that smoothly shifts the random chemical potentials.
During a pump cycle, electrons follow a random walk between resonant orbitals on the chain.
We have shown that winding number $W$ of the random walk around the system equals the Chern number $C$ of the associated electron wavefunction.
This identification leads to some striking predictions, e.g. that generic states in this limit have large, random Chern number.

We have further shown that the above picture is valid away from the thin-torus limit, i.e. for $L_y>1$, as long as the torus aspect ratio $a$ is large enough. 
The crossover between 2D and 1D behavior as $a$ is increased starts at the band center and spreads towards the band edges.
The broadening is predicted to be extremely slow, $\sim \sqrt{\ln(a)}$, but it is nonetheless visible in our numerics at $L_y \sim \mathcal O(10)$, quite far from the thin-torus limit.

On a theoretical level, our findings provide a new example of subtle interplay between topology and disorder~\cite{Prodan2010, Huse2013, Bauer2013, Yasaman2015, Parameswaran2018, Kuno2019}.
The idea of topological pumping, which goes back to Thouless~\cite{Thouless1983}, is a subject of rising theoretical interest, especially in connection to Floquet physics~\cite{Martin2017, Weinberg2017, Kolodrubetz2018, Privitera2018, Wauters2019, Friedman2019} and synthetic dimensions~\cite{Petrides2018}. 
Here it is applied in a new, disordered context, where it provides the key to interpret the quasi-1D limit of the quantum Hall plateau transition.

We conclude with some remarks related to experiment. 
As the non-local avoided crossings that underpin the picture presented here are generally very narrow (exponentially in system size), the adiabatic time scales required to observe this behavior in macroscopic systems are unphysically long. 
However, for microscopic systems, the manipulations required may still be performed adiabatically. 
The ingredients required are 
(i) adiabatically tunable, pseudo-random on-site chemical potentials, 
(ii) nearest-neighbor hopping, and 
(iii) sufficiently long coherence times (relative to the required adiabatic time scale). 
Clean Thouless pumps have been successfully engineered using ultracold bosonic~\cite{Lohse2015, Lohse2018} or fermionic~\cite{Nakajima2015} atoms in optical superlattices, single spins in diamond~\cite{Ma2018}, Bose-Einstein condensates~\cite{Lu2016} and quantum dots~\cite{Switkes1999, Buitelaar2008}; 
adding disorder could be an interesting new direction for these and other experimental platforms.
Finally, while implementing periodic boundary conditions (i.e. arranging the qubits on a circle) in some such platforms may be problematic, the striking coexistence of Anderson localization and non-local charge transport across the length of the one-dimensional quantum simulator would be observable even on an open line segment.

\begin{acknowledgments}
This work was supported by DOE BES grant $\text{DE-SC0002140}$.
We acknowledge useful conversations with Shivaji Sondhi.
\end{acknowledgments}

\bibliography{1dchern}

\clearpage
\widetext


\setcounter{equation}{0}
\setcounter{figure}{0}
\setcounter{table}{0}
\setcounter{page}{1}
\makeatletter
\renewcommand{\thesection}{S\arabic{section}}
\renewcommand{\theequation}{S\arabic{equation}}
\renewcommand{\thefigure}{S\arabic{figure}}

\begin{center}
{\bf Supplemental Material: Dimensional crossover of the integer quantum Hall plateau transition and disordered topological pumping}\\~\\
{Matteo Ippoliti${}^1$ and R. N. Bhatt${}^2$} \\
{\it ${}^1$Department of Physics and ${}^2$Department of Electrical Engineering, Princeton University, Princeton NJ 08544, USA}
\end{center}

\section{Details of numerical calculation of Chern numbers \label{app:numerics}}

We consider a continuum model of electrons in a high magnetic field projected into the LLL. 
The system is set on a rectangular torus with sides $L_x$, $L_y$ and generalized periodic boundary conditions parametrized by angles $\angles \equiv (\theta_x, \theta_y) \in [0,2\pi)^2$.
The torus is pierced by $N_\phi$ quanta of magnetic flux, i.e. $L_x L_y = 2\pi N_\phi$.
The wavefunctions in Landau gauge are given by
\begin{equation}
\psi_{n}(x,y) = \frac{1}{\sqrt{L_y{\pi}^{1/2}} } 
\sum_{p\in \mathbb Z} 
e^{ip \theta_x} e^{i k_{n,p}(\theta_y) y -\frac{1}{2}(x-k_{n,p}(\theta_y))^2}
\label{eq:basis}
\end{equation}
with $k_{n,p}(\theta_y) = 2\pi (n+pN_\phi + \theta_y/2\pi)/L_y$.
Disorder is represented by a Gaussian white noise potential $V_{\mb q}$ obeying 
\begin{equation}
\langle V_{\mb q_1} V_{\mb q_2} \rangle = U^2 \delta^2(\mb q_1+\mb q_2) \;;
\end{equation}
the disorder strength is the only energy scale in the problem and we set it to $U=1$.
The Hamiltonian matrix elements in the basis of \eq{basis} are
\begin{align}
H_{n_1,n_2} & = 
\sum_{p,m_x,m_y\in \mathbb Z} e^{ip\theta_x} \, \delta(n_2-n_1+pN_\phi = m_y) \, \tilde{V}_{\mb q}  \, e^{-2\pi i (n_1+m_y/2+\theta_y/2\pi)m_x/N_\phi}
 \label{eq:hamiltonian}
\end{align}
where $\tilde{V}_{\mb q} \equiv V_{\mb q} e^{-\frac{1}{4}q^2}$ is the LLL-projected potential, and $\mb q = 2\pi (m_x/Lx, m_y/L_y)$.

In order to calculate the Chern number of each eigenstate in the spectrum of $H$ from \eq{hamiltonian}, we follow the standard method described in Ref.~[S1] and split the torus of boundary angles $\angles$ into a lattice, $\angles_{i,j} = 2\pi (i/N_x, j/N_y)$. 
Appropriate values of the mesh size were discussed in Ref.~[S2], which takes $N_x=N_y=\sqrt{4\pi N_\phi/3}$. Unlike the present work, Ref.~[S2] only deals with square systems where $N_x=N_y$ is clearly optimal.  We fix the product to the same value, $N_x N_y \gtrsim 4\pi N_\phi/3$, but we find that a rectangular mesh with $N_x/N_y \simeq L_y/L_x$ is optimal. This is physically reasonable as the angle in the \emph{short} direction affects the \emph{long} boundary, and {\it vice versa}.

For each site on this lattice, we diagonalize \eq{hamiltonian} numerically and obtain the spectrum $\{ |\theta_{ij},n\rangle, n = 0,\dots N_\phi-1 \}$. 
We then assign a $U(1)$ gauge variable to each bond in the lattice: 
\begin{equation}
A_n^x(i,j) \equiv \frac{ \langle \angles_{ij},n | \angles_{i+1,j},n\rangle }{ \left| \langle \angles_{ij},n | \angles_{i+1,j},n\rangle \right| },
\
A_n^y(i,j) \equiv \frac{ \langle \angles_{ij},n | \angles_{i,j+1},n\rangle }{ \left| \langle \angles_{ij},n | \angles_{i,j+1},n\rangle \right| }
\label{eq:gauge_links}
\end{equation} 
for horizontal and vertical bonds, respectively.
The curvature of this gauge field is given by the ``integral'' around a plaquette:
$$
U_n(i,j) \equiv A_n^x(i,j) A_n^y(i+1,j) {A_n^x(i,j+1)}^\ast {A_n^y(i,j)}^\ast \;.
$$
If the mesh is fine enough (i.e. $N_x$, $N_y$ are large enough), $|U_n(i,j)-1| \ll 1$ and one can define $\gamma_n(i,j) \equiv \ln U_n(i,j)$ without ambiguity in the choice of branch cut for the logarithm. 
The Chern number is then 
\begin{equation}
C_n \equiv \frac{1}{2\pi i} \sum_{i,j} \gamma_n(i,j) \;. \label{eq:chern}
\end{equation}
This method also gives us access to all the energies $E_n(i,j)$, which we use to compute the Thouless conductance.

A subtlety implicit in formulae \eq{gauge_links} is that the inner products involve wavefunctions with different boundary conditions. The matrices implementing such inner products are not equal to the identity and must be calculated.
As we sample $\angles$ from a rectangular lattice and only need inner products between neighboring points in that lattice, we are interested in 
\begin{equation}
\mathcal N_{mn} (\angles; \angles + \mu \hat j) \equiv 
\langle \psi_m(\angles) | \psi_n(\angles+\mu\hat j) \rangle
\end{equation}
for $\hat j = \hat x, \hat y$.

For $\hat j=\hat y$, we have
\begin{align}
\mathcal N_{mn} (\angles, \angles+\mu \hat y) 
& = \frac{1}{L_y \sqrt{\pi}} \sum_{P,p \in \mathbb Z}
e^{i p \theta_x} 
\int dy \ e^{i k_{n-m,p}(\mu)y } \int_{PL_x}^{(P+1)L_x} dx\ 
\exp \left[ -\left(x-k_{\frac{n+m}{2}, \frac{p}{2}} (\theta_y+\mu/2)\right)^2 - \frac{1}{4}k^2_{n-m, p}(\mu) \right] \nonumber \\
& = \sum_{p\in \mathbb Z} e^{ip\theta_x} \frac{e^{ik_{n-m,p}(\mu)L_y}-1}{ik_{n-m,p}(\mu)L_y}
e^{-\frac{1}{4} k^2_{n-m,p}(\mu)} \;,
\end{align}
where $k_{n,p}(\mu) = \frac{2\pi}{L_y} (n+N_\phi p + \mu/2\pi)$.
One can verify that this reduces to the identity for $\mu\to 0$. 
Moreover the entries only depend on the difference $n-m$.
For $\hat j = \hat x$, we have instead
\begin{align}
\mathcal N_{mn} (\angles, \angles+\mu \hat x) 
& = \frac{1}{L_y \sqrt{\pi}} \sum_{P,p \in \mathbb Z}
e^{i p \theta_x + i(p+P)\mu} 
\int dy \ e^{i k_{n-m,p}(0)y} \int_{0}^{L_x} dx\ 
\exp \left[ -\left(x-k_{\frac{n+m}{2}, P+\frac{p}{2}}(\theta_y)\right)^2 - \frac{1}{4}k^2_{n-m, p}(0) \right]  \nonumber \\
& =\delta_{mn} \frac{1}{\sqrt{\pi}} \sum_{P\in \mathbb Z} 
e^{iP\mu} \int_{PL_x}^{(P+1)L_x} dx\ e^{-(x-k_{n,0}(\theta_y))^2} \;,
\end{align}
where we used the fact that the $y$ integral gives a $\delta$ function to arrive to the result.
This matrix is diagonal and reduces to the identity for $\mu \to 0$ as well.
Moreover, we verify numerically that for $L_x=L_y$ the two matrices are conjugate to each other.
This reflects the fact that the two must map into each other under a $\pi/2$ rotation followed by a gauge transformation.


\section{Additional data on density of Chern states}
\begin{figure}[h!]
\centering
\includegraphics[width=0.9\textwidth]{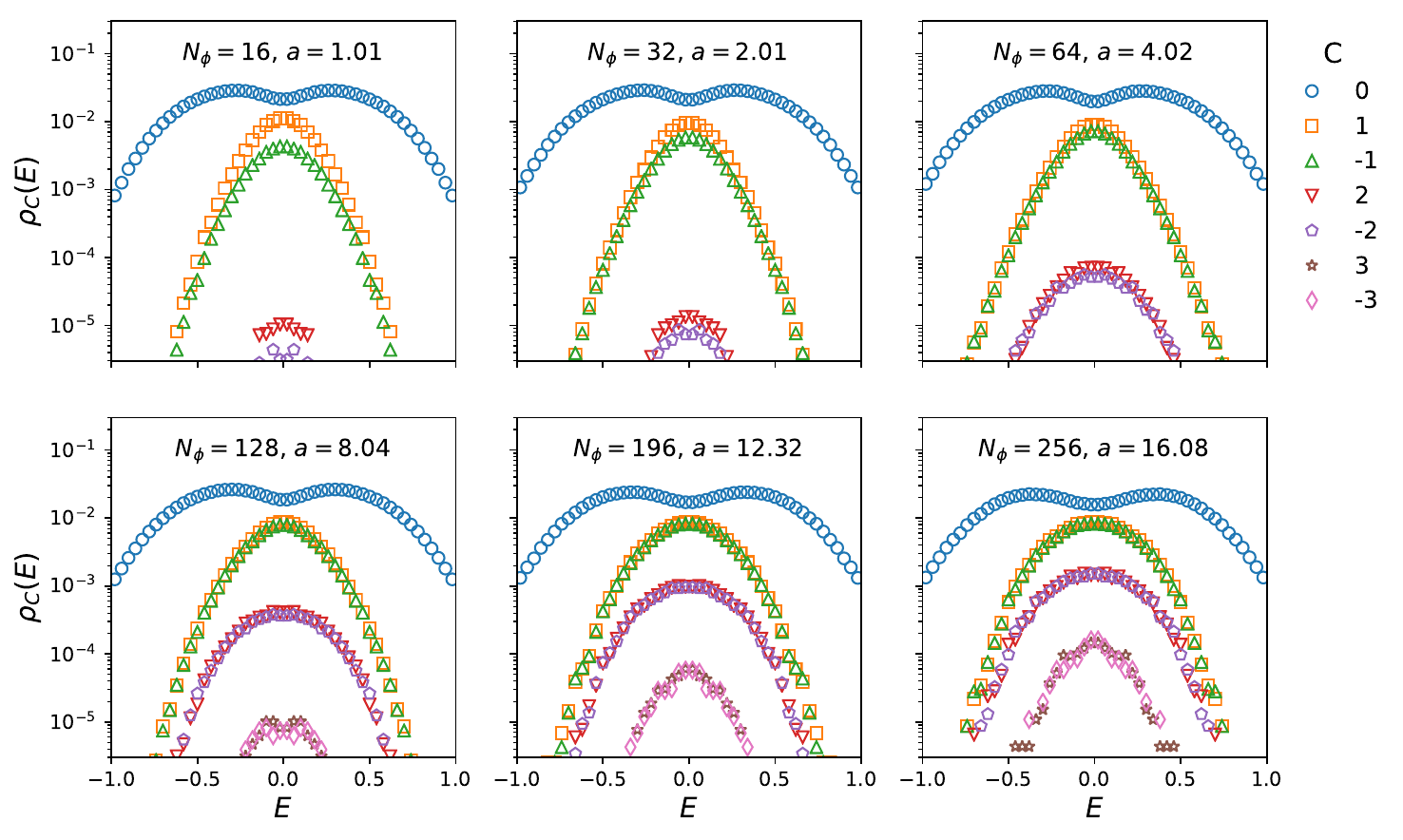}
\caption{Density of states with Chern number $C$, $\rho_C(E)$, for rectangular tori with $L_x=10\ell_B$ and variable $N_\phi$ indicated on each panel. 
As $N_\phi$ increases the system gradually approaches the 1D regime: more and more states become topological ($C\neq 0$), due both to the broadening of $\rho_{\pm 1}(E)$ and to states with higher $|C|$ becoming more common at the band center. \label{fig:cherndata}}
\end{figure}

\begin{figure}[h!]
\centering
\includegraphics[width=0.7\textwidth]{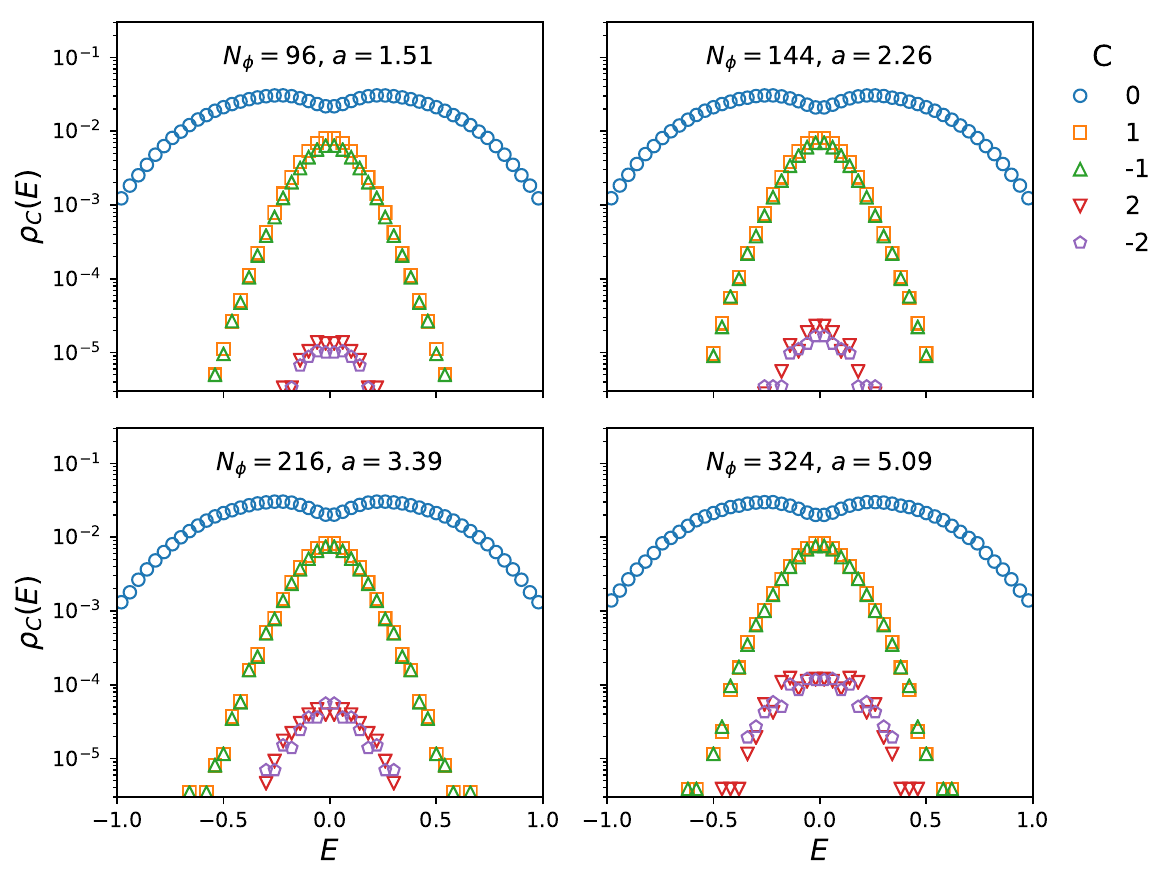}
\caption{Same as \fig{cherndata}, but for $L_x=20\ell_B$ instead of $10\ell_B$. 
Despite the larger system size $N_\phi$, the aspect ratio we can achieve is smaller and the system is further away from the 1D limit: $C=\pm 3$ states are still uncommon (below the displayed range) and the broadening of $\rho_{\pm 1}$ is not as clear. Nonetheless the growth of $C = \pm 2$ states is visible.
\label{fig:cherndata20}}
\end{figure}

Here we report additional data on the density of states $\rho_C(E)$ parsed by Chern number $C$ to complement the results in Fig.~1 in the main text.
We consider rectangular tori with $L_x = 10\ell_B$ (\fig{cherndata}) and $L_x = 20\ell_B$ (\fig{cherndata20}) and scale $L_y$.
Doing so causes both a broadening of $\rho_{\pm 1}$ and an increase in the density of higher-$|C|$ states.
Comparing the two $L_x$ values shows how this behavior is due to the 2D-1D crossover controlled by the aspect ratio $a = L_y/L_x$, as opposed to 2D behavior controlled solely by the system size $N_\phi$.

\begin{figure}[h!]
\centering
\includegraphics[width=0.9\textwidth]{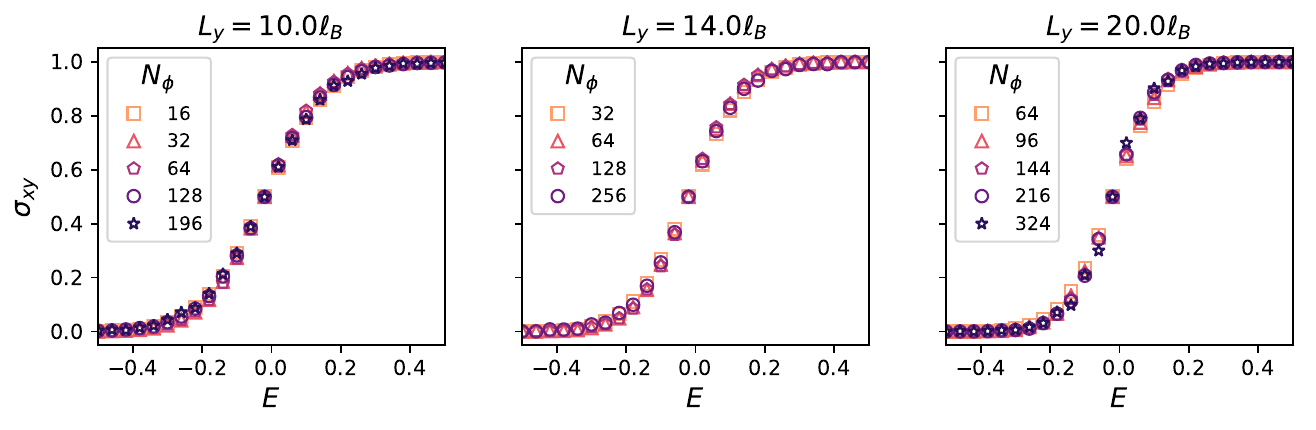}
\caption{Disorder-averaged Hall conductance $\sigma_{xy}$ calculated from the densities $\rho_C$ as in \eq{sigma}.  \label{fig:sigma}}
\end{figure}

While the density of Chern-$C$ states broadens for each $C$, the linear combination $\sum_C C\rho_C$ remains constant, as seen by looking at the disorder-averaged Hall conductance,
\begin{equation}
 \sigma_{xy}(E) = \int_{-\infty}^E d\varepsilon\ \sum_{C\in\mathbb Z} C\rho_C(E) \;,
 \label{eq:sigma}
\end{equation}
shown in \fig{sigma}.
Under 2D scaling, $\sigma_{xy}(E)$ is known to collapse onto $F(E L^{1/\critexp})$, with $F$ a scaling function.
We conclude that the width of the crossover interval between the two conductance plateaus is fixed by the smallest side of the torus.

\pagebreak


\section{Proof of the identity between Chern number and winding number}

\begin{figure}
\centering
\includegraphics[width=0.6\textwidth]{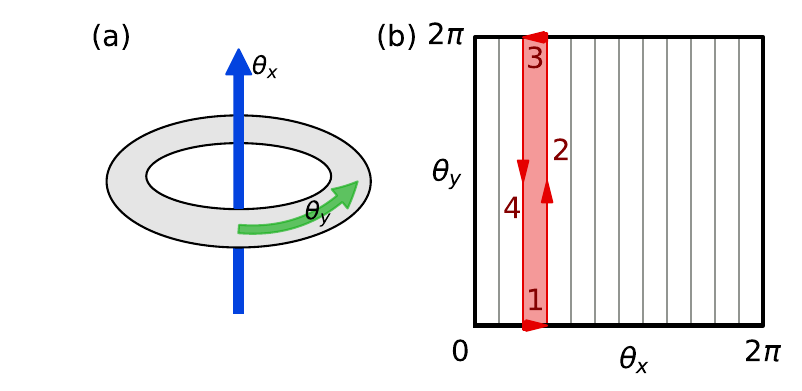}
\caption{\label{fig:scc}
(a) Sketch of a thin torus system. The two boundary angles correspond to fluxes through the long ($\theta_x$) and short ($\theta_y$) non-trivial loops of the torus, respectively.
(b) The partitioning of the $\angles$ torus used to calculate the Chern number.
All slices contribute the same amount. The sides of the highlighted slice are numbered according to the discussion in the text.}
\end{figure}

\begin{figure*}
\centering
\includegraphics[width=\textwidth]{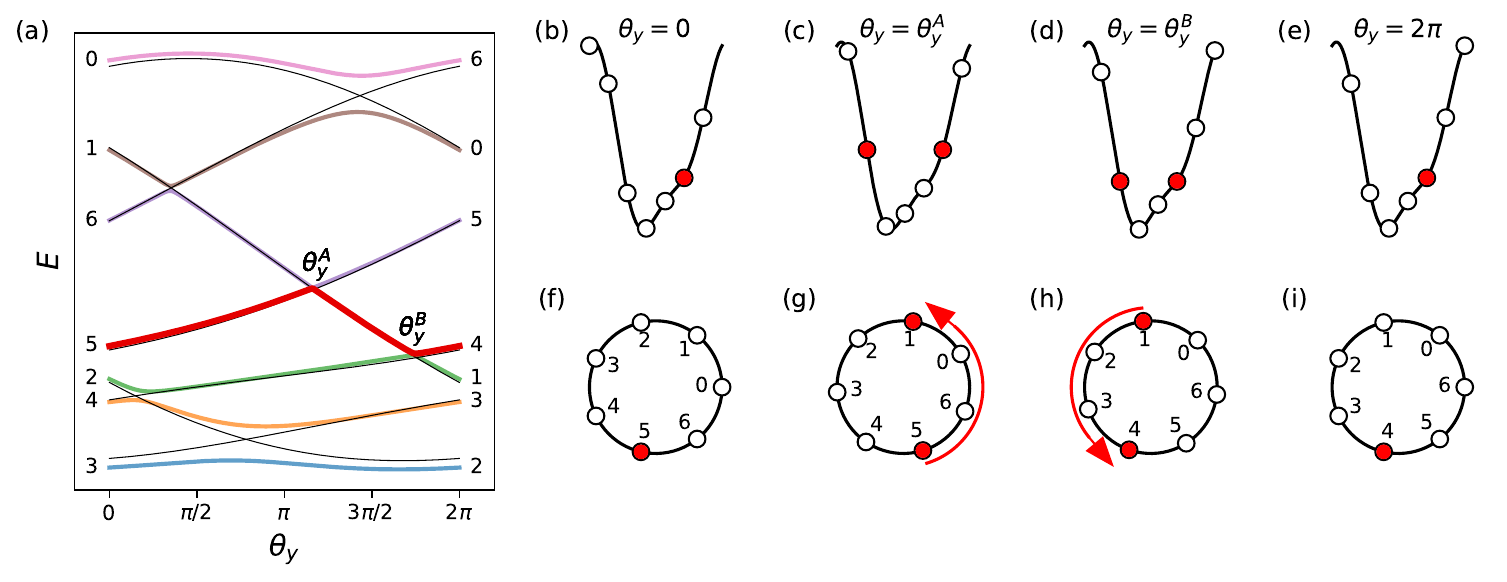}
\caption{(a) Energy levels (thick lines) and on-site chemical potentials (thin lines) as a function of Thouless pump parameter $\theta_y$ for a simple example with $N=7$ orbitals. 
All states have $C=0$ except the one starting at orbital 5. 
(b-e) Evolution of the chemical potentials at four points during the cycle, $\theta_y=0$, $\theta_x^A$, $\theta_y^B$, $2\pi$. White circles denote empty orbitals, red circles occupied orbitals. At $\theta_y^{A,B}$ the occupied orbital is at resonance with an empty orbital, an avoided crossing opens, and the electron hops.
(f-i) View of the same cycle in real space, showing how the electron's path winds once around the system during each cycle. \label{fig:network}}
\end{figure*}

We calculate the Chern number by splitting the ``Brillouin zone'' $\angles \in [0,2\pi)^2$ in many vertical strips, $\theta_y \in [0,2\pi)$, $\theta_x \in [m\varepsilon, (m+1)\varepsilon)$, for $\varepsilon = 2\pi/M\ll 1$, $m,M \in \mathbb Z$, as sketched in Fig.~\ref{fig:scc}.
Similarly to the numerical approach, 
we calculate the phase for transporting the wavefunction around each side of each small rectangle, then add all the contributions. The cycle goes as follows:
\begin{enumerate}
\item Increase $\theta_x$ from $m\varepsilon$ to $(m+1)\varepsilon$ at fixed $\theta_y$.
This twist of boundary conditions has no effect on a perfectly Anderson-localized orbital.
\item Increase $\theta_y$ from 0 to $2\pi$ at fixed $\theta_x$. 
During this step, the electron winds $W_i$ times around the circle and acquires an Aharonov-Bohm phase $e^{i\theta_x W_i} = e^{i (m+1)\varepsilon W_i}$, where $\theta_x = (m+1)\varepsilon$ is the flux through the circle. An example is sketched in \fig{network}
\item Decrease $\theta_x$ from $(m+1)\varepsilon$ to $m\varepsilon$. Like step 1, this has no effect.
\item Decrease $\theta_y$ from $2\pi$ to 0. The electron winds $-W_i$ times around the circle (as the evolution is perfectly reversible) and acquires a phase $e^{-i W_im\varepsilon}$, completing the cycle. 
\end{enumerate}
The net phase acquired by the electron wavefunction is $U_m = e^{i W_i \varepsilon}$. 
Since $\varepsilon \ll 1$, one can take the logarithm unambiguously and obtain the Chern number:
\begin{equation}
C_i = \frac{1}{2\pi i } \sum_{m=1}^M \ln U_m =  W_i \;.
\label{eq:CtoW}
\end{equation}

This derivation relies on a few assumptions, namely that (i) the electron orbital is Anderson-localized at every $\theta_x = m\varepsilon$, and (ii) $W_i$ does not depend on $\theta_x$.
The first assumption is satisfied in the thin-torus limit. The electron typically resonates between orbitals at a distance $\sim \mathcal O(N_\phi)$, so the gap at the avoided crossing is $\sim (t/v)^{N_\phi}$. 
Therefore the fraction of the interval $\theta_x \in [0,2\pi)$ spent in a non-local superposition is exponentially small in system size and asymptotically has measure 0. 
That $W_i$ doesn't depend on $\theta_x$ in this limit follows from similar considerations.

A natural consistency check on the identification of Chern number and winding number is that one should always have $\sum_i W_i = 1$, so as to match the total Chern number of the Landau level. 
Avoided crossings (absent fine-tuning) happen at isolated points in the $\theta_x$ interval and involve only a pair of states each.
Therefore, the question can be rephrased as a simple combinatorial problem: 
showing that any decomposition of the cyclic permutation $\mathcal C : (1,2,\dots N) \mapsto (2,3,\cdots N, 1)$ (representing the spectral flow) as a product of pairwise swaps $\mathcal S_{ij}$ (representing the avoided crossings) gives a total winding number of 1. 

This can be shown as follows. Let one such decomposition be $\mathcal C = \prod_c \mathcal S_{n_1(c),n_2(c)}$. 
Define a matrix of integers $d_{c,i}$ as the distance element $i$ traverses during move $c$, with sign (this is 0 for all elements except the two being swapped, $i=n_{1,2}(c)$, which have opposite values). 
By this definition, the sum of each row is trivially $\sum_i d_{c,i} = 0$.
The sum of each column $i$ is the total distance traversed by element $i$, which must be $-1 \mod N$; we set $\sum_c d_{c,i} = NW_i - 1$, defining the winding number $W_i$.
Putting the two together, we get
\begin{equation}
\sum_{c,i} d_{c,i} =  N\left(-1+\sum_i W_i\right) = 0
\end{equation}
hence $\sum_i W_i = 1$.
In a Chern-$k$ band, one would have to replace $\mathcal C$ with $\mathcal C^k$ (as the spectral flow cycles the orbitals $k$ times in a Chern-$k$ band) and correctly obtain $\sum_i W_i = k$.


\section{Random walk of electrons during a pump cycle}

Here we discuss details of the random walk that electrons in the system undergo during a Thouless pump cycle as a result of resonances between the random on-site potentials.
In particular, we justify the following statements in the main text:
(i) that electrons move along the shortest path between resonant orbitals (even when taking randomness into account), and (ii) that the random walks become infinitely long in the thin torus limit, leading to large, random real-space winding numbers.

\subsection{Electrons' path between resonant sites}

When two sites $n_1$ and $n_2$ are near resonance at some point during the pump cycle, charge transport between them always happens through the \emph{shortest} path.
This fact is crucial for the definition of the winding number.

To address this point, we must fist justify the use of the truncated Hamiltonian, where hopping terms with range greater than 1 have been dropped:
\begin{equation}
H_{1D} = \sum_n v_n c^\dagger_n c_n + (t_n c_n^\dagger c_{n+1} + h.c.) \;.
\label{eq:1dham}
\end{equation}
Such longer-range hopping terms, which are generically present, are induced by higher wave vectors $q_y = 2\pi d/L_y$  ($\mathbb Z \ni |d|>1$ is the hopping range) in the random potential $V(\mb q)$.
As such, they are suppressed as $\sim e^{-d^2 \pi^2 /L_y^2}$ by LL projection.
In contrast, the nearest-neighbor hopping term $t\sim e^{-\pi^2/L_y^2}$ can transfer an electron over the same distance at order $d$ in perturbation theory, with amplitude $\sim (t/v)^d = \mc O(e^{-d \pi^2/L_y^2})$. 
This is asymptotically dominant over any processes involving hopping with range longer than 1 in the $L_y\to0$ limit, thus justifying the use of the truncated Hamiltonian in \eq{1dham}.

Having shown that the leading amplitudes come from perturbation theory in the nearest neighbor hopping $t$, we can estimate the amplitudes of the two paths (clockwise and counterclockwise) connecting sites $n_1$ and $n_2$:
\begin{equation}
|A_\pm| \sim t \prod_{n \in \mc P_\pm } \frac{t}{|v_n - v_{n\pm 1}|} \equiv t e^{-w(\mc P_\pm)} \;,
\end{equation}
where $\mc P_i$ labels the path ($\mc P_+ = \{n_1, n_1+1,\cdots n_2-1, n_2\}$ is clockwise, $\mc P_- \{ n_1, n_1-1, \cdots n_2+1, n_2\}$ counterclockwise).
The two paths have lengths $d_+ \equiv n_2-n_1$ (we assume $n_2>n_1$ without loss of generality) and $d_- \equiv N_\phi - d_+$. 
Which one is favored depends on the ratio of the amplitudes, 
\begin{equation}
\left| \frac{A_-}{A_+} \right|  = \exp\left[ w(\mc P_+) - w(\mc P_-) \right] \;.
\end{equation}
The quantity in the exponent is a random variable, as it depends on the random chemical potentials $v_n$:
\begin{equation}
w(\mc P_\pm) = \sum_{n\in \mc P_\pm} \log \frac{|v_n - v_{n\pm 1} |}{t} \;.
\label{eq:paths_amplitude}
\end{equation}
For the case of interest, the $v_n$ are uncorrelated normal variables of zero mean and variance $U^2$ (the spatial separation between Landau orbits is $2\pi/L_y \gg 1$ in the thin-torus limit, hence correlations between different $v_n$ can be neglected).
The mean and variance of the quantities $w(\mc P_\pm)$ can be calculated exactly:
\begin{equation}
\mathbb E[w(\mc P_\pm)] = d_\pm \left( \ln \frac{U}{t} - \frac{1}{2} \gamma \right) \;,
\qquad
\text{Var}[w(\mc P_\pm)] = d_\pm \frac{\pi^2}{8}
\end{equation}
where $\gamma = 0.577\dots$ is the Euler-Mascheroni constant.
For long paths ($d_\pm \gg 1$) the exponent of \eq{paths_amplitude} becomes sharply peaked around a mean value $\mu \equiv (d_+ - d_-) \ln(e^{-\gamma/2}U/t) \propto (d_+ - N_\phi/2)$, with a sub-extensive standard deviation $\sigma = (\pi/2\sqrt{2}) \sqrt{d_+ + d_-} \propto N_\phi^{1/2}$. 
Therefore the shorter path is exponentially favored over the longer one, unless the two sites $n_1$, $n_2$ are diametrically opposite on the ring ($d_+  = \frac{N_\phi}{2}$) within a relative error $\sim N_\phi^{-1/2}$.
This happens with zero probability in the thermodynamic limit, and shows that, even when taking disorder into account, electrons almost always (i.e. up to zero-probability exceptions) follow the shortest path between resonant orbitals.

\subsection{Length of the random walks}

The sequence of resonances encountered by an electron during a pump cycle determines a random walk whose winding number coincides with the Chern number. 
Here we aim to show that the typical length of these random walks diverges in the 1D thermodynamic limit.

In the thin-torus limit, the Hamiltonian becomes approximately diagonal, $H_{1D} \simeq \sum_n v_n(\theta_y) c_n^\dagger c_n$, with the $v_n$ essentially uncorrelated normal variables with zero mean and variance $U^2=1$.
This is because $v_n = V_0(x_n)$, and the ``sites'' $x_n = (2\pi n + \theta_y)/L_y$ are separated by a distance $2\pi/L_y \gg 1$, whereas the LL-projected random potential $V_0(x)$ is correlated over a length $\mc O(1)$.
The task of determining the length of the random walk therefore amounts to counting the number of resonances $N_r$ between the functions $\{ v_n(\theta_y) \}$ as $\theta_y$ is tuned through a cycle.
The number of times two such functions $v_{1}(\theta_y)$, $v_2(\theta_y)$ cross is $\propto L_y^{-1}$ (i.e. directly proportional to the inter-orbital distance $2\pi/L_y$).
The total number of resonances between any two orbitals is thus $\sim \binom{N_\phi}{2}/L_y$. 
Finally, the number of resonances encountered by a given electron is $N_r \sim N_\phi/L_y \sim L_x$.

We thus predict that each electron undergoes a random walk with a divergent number of steps $N_r \propto L_x$. 
The winding number is
\begin{equation}
W \equiv \frac{1}{N_\phi} \left(1+ \sum_i d_i\right)\;,
\label{eq:W_from_d}
\end{equation} 
Approximating each step $d_i$ as a uniformly distributed variable in $[-N_\phi/2, N_\phi/2]$, 
the variance of $W$ is
\begin{equation}
\text{Var}[W]
 = \frac{1}{N_\phi^2} \sum_{i=1}^{N_r} \mathbb E[d_i^2]
 = \frac{N_r}{12} \;.
\end{equation}
From this we conclude that the typical winding number is $W_{\rm typ} \sim L_x^{1/2}$.

We can corroborate the above analysis with numerical simulations.
We do so by generating random realizations of $V_0(x)$ and, for each one, counting the resonances one electron encounters during the pump cycle.
In practice, at $\theta_y=0$, we define a permutation $e\in S_{N_\phi}$ that sorts the on-site chemical potentials: 
$$
v_{e(1)}(0) < v_{e(2)}(0) < \dots < v_{e(N_\phi)}(0) \;.
$$
As $\theta_y$ is increased, whenever we encounter a resonance $v_{e(k)}(\theta^\star) = v_{e(k+1)}(\theta^\star)$, we update the permutation to reflect the reordering of orbitals:
$e\mapsto e'$ with $e'(k) = e(k+1)$, $e'(k+1) = e(k)$, $e'(j) = e(j)$ otherwise.
We increment $\theta_y$ in small finite steps until two entries in $e$ change value.
If more than two entries are found to change over one step, we undo the step, halve the step size, and repeat until we can resolve the process unambiguously into elementary swaps.

When the cycle is over, each $e(k)$ must return to its initial value minus one, as $v_{n-1}(2\pi) = v_n (0)$ for all $n$.
The $e(k)$ so defined keeps track of the real-space position of the electron with $k^{\rm th}$ highest energy throughout the cycle.
For each electron $k$, we can thus extract the chronologically ordered list of positions $n_1, n_2, \dots n_{N_r+1}$ occupied during the cycle. 
$N_r$ is simply the number of resonances encountered by electron $k$.
Spectral flow implies $n_{N_r+1} = n_1- 1$.
To calculate the winding number, we define
\begin{equation}
d_i \equiv \left( n_{i+1} - n_i + \frac{N_\phi}{2} \text{ mod } N_\phi \right) - \frac{N_\phi}{2}
\label{eq:d_def}
\end{equation}
($|d_i| \leq N_\phi/2$ is the length of the shortest path between $n_i$ and $n_{i+1}$ and the sign of $d_i$ specifies its direction, clockwise or counterclockwise), which can be plugged into the definition \eq{W_from_d}.

\begin{figure}
\centering
\includegraphics[width=0.8\textwidth]{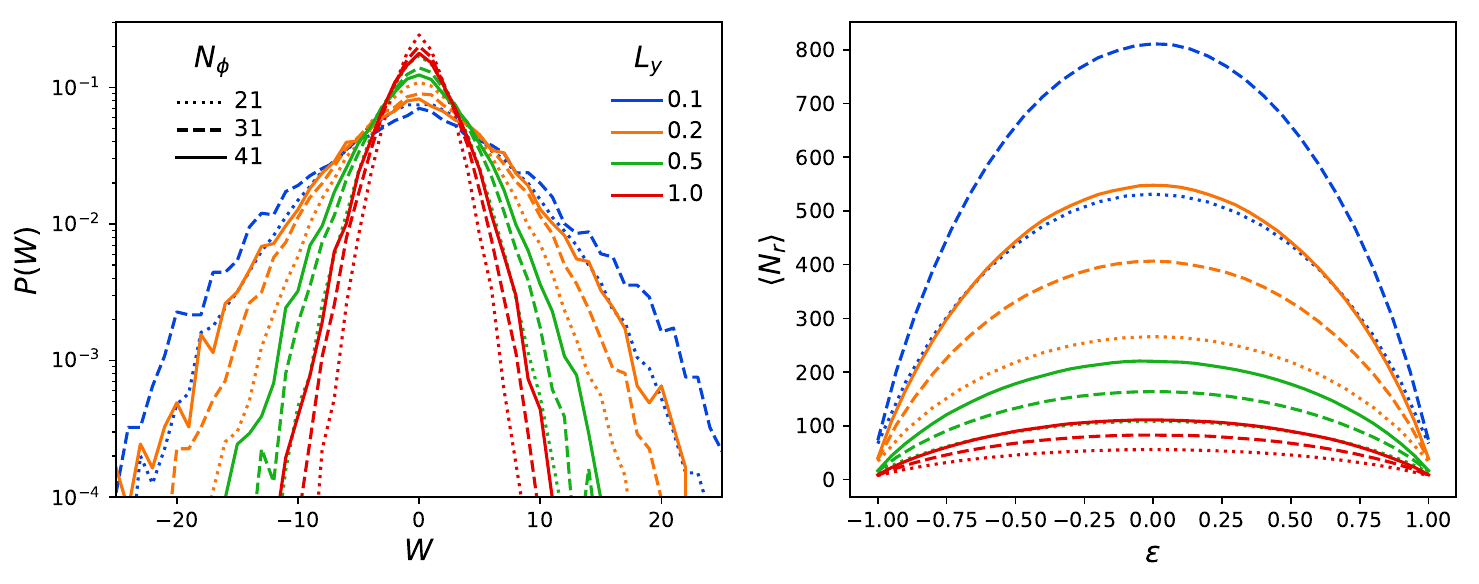} \\
\includegraphics[width=0.8\textwidth]{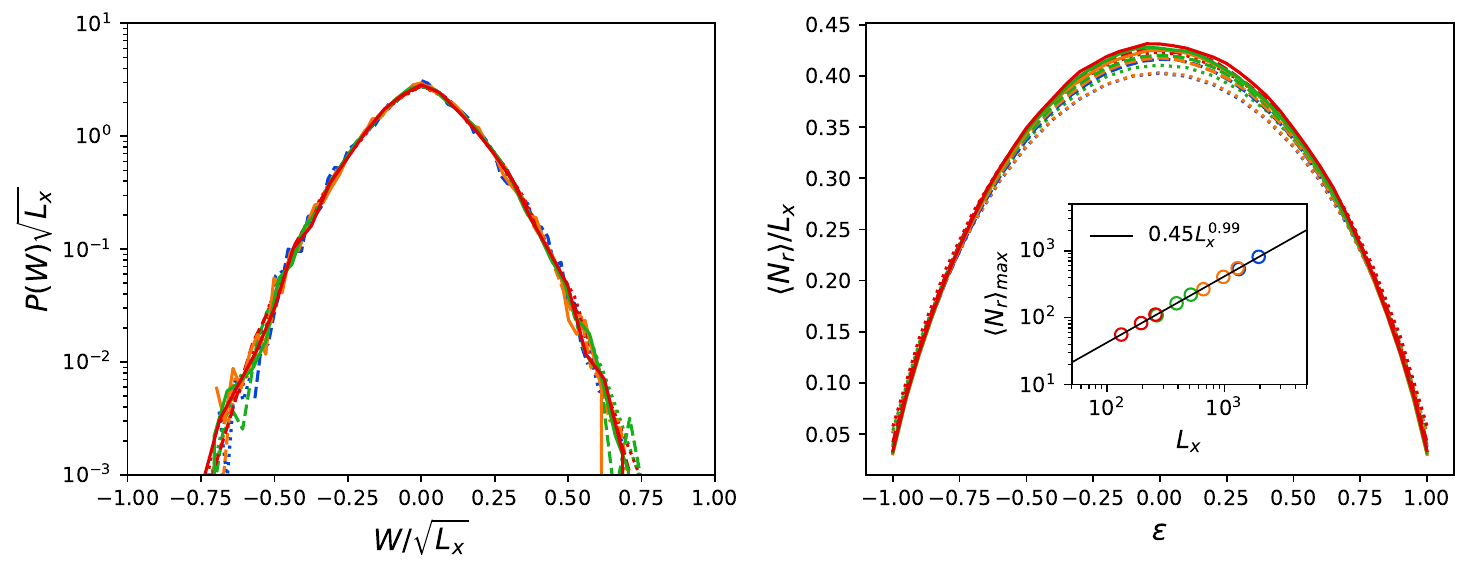}
\caption{Numerically obtained statistics of random walks.
Top: probability distribution of winding number $W$ (left) and average number of resonances encountered during a cycle $\langle N_r\rangle$ as a function of normalized energy $\epsilon$ (right), for various numbers of flux quanta $N_\phi$ and lengths of the short circumference $L_y$.
Bottom: same quantities collapsed by rescaling the axes with appropriate powers of $L_x = 2\pi N_\phi / L_y$ (long circumference). Bottom right inset: scaling of the maximum of $\langle N_r \rangle$ with $L_x$. The best fit yields $\langle N_r \rangle \propto L_x^{0.99}$, in good agreement with the linear prediction.
We average over $10^4$ potential realizations at the smallest size and 500 at the largest.
\label{fig:rw}}
\end{figure}

In Fig.~\ref{fig:rw} we show data on the statistics of the random walk length $N_r$ and winding number $W$ for varying system sizes and shapes. 
The average of $N_r$ is plotted as a function of the normalized electron energy $\epsilon \equiv 2k/N_\phi-1$ ($k$ again enumerates the electrons from lowest to highest energy, so $-1\leq \epsilon \leq 1$, with $\epsilon=0$ the band center).
We see that $\langle N_r \rangle \simeq L_x f(\epsilon)$, with $f(\epsilon)$ a function of energy, confirming our analytical derivation.
$f(\epsilon)$ is even (due to statistical particle-hole symmetry of the disorder),
has a maximum at the band center, 
is monotonically decreasing towards the band edges, 
and is positive for all $-1< \epsilon < 1$.
Our expectation of diffusive scaling for the winding number, i.e. of a distribution
$$
P(W) = \frac{1}{\sqrt{2\pi\sigma^2}} e^{-\frac{1}{2}(W/\sigma)^2} \;, \qquad 
\sigma \sim N_r^{1/2} \sim L_x^{1/2} \;,
$$
is also borne out by numerics, as can be seen from the collapse of the distributions $P(W/\sqrt{L_x})$ (also in Fig.~\ref{fig:rw}).

These data confirm that in the extreme thin-torus limit {almost all} electrons acquire a large, random Chern number. 
Exceptions appear to be possible only for the absolute highest and lowest energies in the band ($\epsilon = \pm 1$), involving a sub-extensive number of states.


\section{Electron localization length}

To complement our analysis of the localization properties of the system, we calculate the electron localization length in the long direction.
We consider the real-space density profile of an eigenstate $\alpha$ integrated over the short direction,
\begin{equation}
\bar{\rho}_\alpha (x) \equiv \int_0^{L_y} dy\, | \psi_\alpha (x,y) |^2 \;.
\end{equation}
We calculate the localization length $\xi$ by means of the inverse participation ratio, 
\begin{equation}
\textsf{IPR}_{\alpha} \equiv \int_0^{L_x} dx\ ( \bar{\rho}_\alpha(x)) ^2 \;,
\label{eq:iprdef}
\end{equation}
which equals $1/L_x$ for a maximally extended state $\bar{\rho}(x) = {\text const.}$ and $(2\xi)^{-1}$ for an exponentially localized state $\bar{\rho}(x) \propto e^{-2|x|/\xi}$.
We thus define $\xi_\alpha \equiv (2 \textsf{IPR}_\alpha)^{-1}$.
With this definition, we have $\xi_0 \leq \xi \leq L_x/2$, where the lower bound $\xi_0 \equiv \sqrt{\pi/2} \simeq 1.25$ is attained by a single Landau orbital.

The integral \eq{iprdef} can be simplified numerically by decomposing $\psi_\alpha$ in the Landau orbit basis:
\begin{equation}
\bar{\rho}_\alpha (x) =  \frac{1}{\sqrt \pi} \sum_{n=1}^{N_\phi} | \langle n \ket{ \psi_\alpha} |^2 \sum_{p\in \mathbb Z} e^{-(x-pL_x)^2}  \;.
\end{equation}
Performing the Gaussian integrals gives
\begin{equation}
\textsf{IPR}_\alpha =  \sum_{n_1, n_2} 
|\langle n_1 \ket{\psi_\alpha} |^2   |\langle n_2 \ket{\psi_\alpha} |^2  F_{n_1-n_2} \;,
\qquad
F_n \equiv \frac{1}{\sqrt{2\pi}} \sum_{p\in \mathbb Z} e^{-\frac{2\pi^2}{L_y^2} (n-pN_\phi)^2 } \;,
\end{equation}
which can be calculated directly from the basis decomposition of the eigenstates without having to perform a real-space integral numerically;
additionally, the expression for $\textsf{IPR}_\alpha$ can be calculated very efficiently (in time $\mc{O}(N_\phi \ln N_\phi)$) by using the fast Fourier transform.
We also note that in the thin torus limit ($L_y \to 0$) we have $F_n \propto \delta_{n,0}$ and therefore $\textsf{IPR}_\alpha = \sum_n |\langle n \ket{\psi_\alpha}|^4$, which is the expression for a 1D chain with site orbitals $\ket{n}$.

\begin{figure}
\centering
\includegraphics[width=0.8\textwidth]{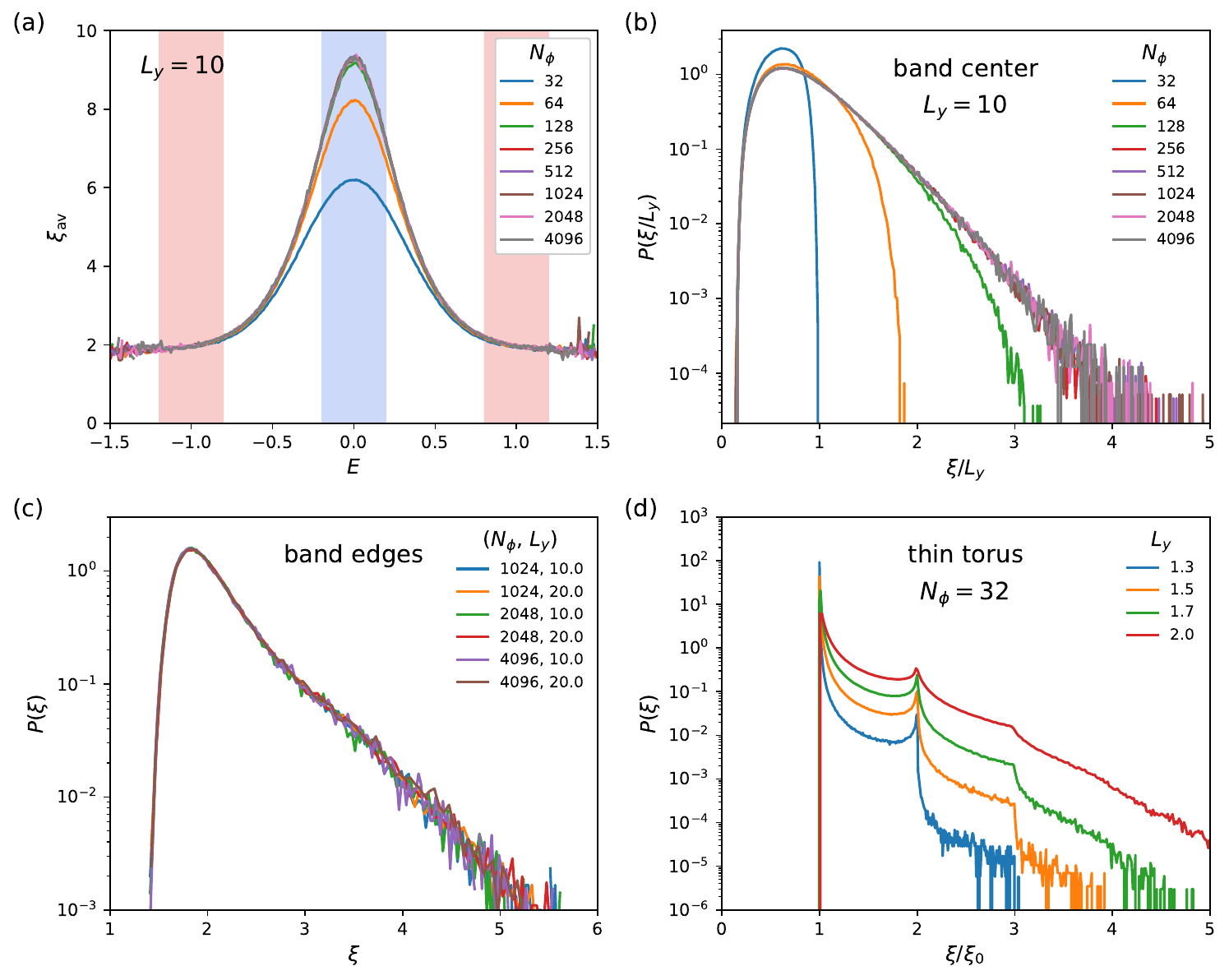}
\caption{
Numerical results for the electron localization length.
The data are averaged over a number of realizations between $10^6$ (for $N_\phi \leq 64$) and $10^3$ (for $N_\phi = 4096$).
(a) Sample- and eigenstate-averaged localization length, $\xi_{\rm av}$, as a function of energy for fixed $L_y = 10$ and increasing $N_\phi$. 
$\xi_{\rm av}$ is size-independent at the band edges, while at the band center it increases up to a saturation value close to $L_y$. 
(b) Distribution of $\xi/L_y$ at the band center (blue energy window in (a)) for fixed $L_y=10$ and increasing $N_\phi$. 
At small sizes, the upper bound $\xi<L_x/2$ is visible. 
At larger sizes, the distribution develops an exponential tail.
(c) Distribution of $\xi$ at the band edges (red energy windows in (a)) for several values of $L_y$ and $N_\phi$. The distribution is essentially size-independent and peaked slightly above the lower bound $\xi_0 = \sqrt{\pi/2} \simeq 1.25$.
(d) Similar data for small systems ($N_\phi = 32$) approaching the thin torus limit.
Singular features at integer multiples of the single-site localization length $\xi_0$ become clearly visible.
\label{fig:loclen}}
\end{figure}

Without the stability issues of the Chern number calculation, we can obtain data for the localization length up to much larger sizes and aspect ratios. 
Results are shown in Fig.~\ref{fig:loclen}.
We find that the average localization length $\xi_{\rm av}(E)$ is peaked at the band center with a finite maximum value $\xi_{\rm av}(0) \propto L_y$ set by the short side. 
(Notice we are measuring $\xi$ in the long direction, so this is fact does not automatically follow from the definition).
At the band edges, the distribution $P(\xi)$ is peaked at an $\mc{O}(1)$ value and nearly independent of $L_x$ or $L_y$. 
On the other hand, at the band center, we find $P(\xi)$ in fact depends (in the $L_x\gg 1$ limit) only on the ratio $\xi/L_y$.
The distribution has an exponential tail, 
$$
P(\xi/L_y) \propto e^{-c \xi/L_y} \;,
$$
with $c \simeq 3$.
This confirms the existence of exponentially rare states with abnormally high sensitivity to changes in $\theta_x$ (boundary angle in the long direction) which may explain the appearance of the 2D critical exponent in the Thouless conductance. 

Finally, we calculate the distribution of $\xi$ for very thin tori (Fig.~\ref{fig:loclen}(d)), where the inter-orbital separation begins to exceed a magnetic length.
There we observe that, while the exponential tail persists, additional structure emerges
at integer multiples of the single-site localization length $\xi_0$~[S3]. 
Values $\xi_0 \leq \xi \leq 2\xi_0$ correspond to weight being shared in some proportion between two orbitals ($\xi = \xi_0$ has all the weight in one orbital, $\xi=2\xi_0$ is equally split between the two); 
values $2 \xi_0 \leq \xi \leq 3\xi_0$ require weight being shared between three orbitals, which requires an additional fine-tuned near-degeneracy; {\it etc}. 
As a result, $P(\xi)$ exhibits ``plateaus'' (or rather slow decay) for $n<\xi/\xi_0<n+1$, and sharp, singular decay at $\xi/\xi_0 = n$ (where $n$ is any integer).  
This confirms the picture of the thin-torus limit being characterized by fully Anderson-localized states, with rare 2-site resonances (and negligible resonances among 3 or more sites).

\vspace{1cm}
{\bf \large References}
\begin{enumerate}[label={[S\arabic*]}]
\item  T. Fukui, Y. Hatsugai, and H. Suzuki, Journal of the
Physical Society of Japan {\bf 74}, 1674 (2005).
\item Q. Zhu, P. Wu, R. N. Bhatt, and X. Wan, Phys. Rev. B {\bf 99}, 024205 (2019).
\item Such ultraviolet ``lattice commensuration'' effects have been previously seen in states near the band edges of the one-dimensional Anderson model at large disorder.
See Sonika Johri, Ph.D. thesis, Princeton University (2015) and Sonika Johri and R. N. Bhatt (in preparation).
\end{enumerate}

\end{document}